\documentclass[12pt]{iopart}

\usepackage[dvips]{graphicx}
\usepackage{amssymb}
\usepackage{amsfonts}

\begin{document}

%%%%%%%%%%%%%%%%%%%%%%%%%%%%%%

\title{R\'enyi entropies as a measure of the complexity of counting
  problems}

\author{Claudio Chamon$^1$ and Eduardo R Mucciolo$^2$}
\ead{\mailto{chamon@bu.edu} and \mailto{mucciolo@physics.ucf.edu}}

\address{$^1$ Department of Physics, Boston University, Boston, MA
  02215, USA}

\address{$^2$ Department of Physics, University of Central Florida,
  Orlando, FL 32816, USA }

%\maketitle

%%%%%%%%%%%%%%%%%%%%%%%%%%%%%%%%%%%%%%%%%%%%%%%%%%%%%%%%%%%%%%%%

\begin{abstract} 
Counting problems such as determining how many bit strings satisfy a
given Boolean logic formula are notoriously hard. In many cases, even
getting an approximate count is difficult. Here we propose that
entanglement, a common concept in quantum information theory, may
serve as a telltale of the difficulty of counting exactly or
approximately. We quantify entanglement by using R\'enyi entropies
$S^{(q)}$, which we define by bipartitioning the logic variables of a
generic satisfiability problem. We conjecture that $S^{(q\rightarrow
  0)}$ provides information about the difficulty of counting solutions
exactly, while $S^{(q>0)}$ indicates the possibility of doing an
efficient approximate counting. We test this conjecture by employing a
matrix computing scheme to numerically solve \#2SAT problems for a
large number of uniformly distributed instances. We find that all
R\'enyi entropies scale linearly with the number of variables in the
case of the \#2SAT problem; this is consistent with the fact that
neither exact nor approximate efficient algorithms are known for this
problem. However, for the negated (disjunctive) form of the problem,
$S^{(q\rightarrow 0)}$ scales linearly while $S^{(q>0)}$ tends to zero
when the number of variables is large. These results are consistent
with the existence of fully polynomial-time randomized approximate
algorithms for counting solutions of disjunctive normal forms and
suggests that efficient algorithms for the conjunctive normal form may
not exist.
\end{abstract}

\pacs{89.70.Eg, 89.70.Cf, 03.67.Mn}
% 02.70.Rr: General statistical methods
% 02.70.-c: Computational techniques; simulations
% 02.50.-r: Probability theory, stochastic processes, and statistics
% 03.67.Mn: Entanglement measures, witnesses, and other characterizations
% 89.70.Cf: Entropy and other measures of information
% 89.70.Eg: Computational complexity

\ams{03D15, 68Q15}
% 03D15: Complexity of computation
% 68Q15: Complexity classes (hierarchies, relations among complexity classes)
% 68Q25: Analysis of algorithms and problem complexity

\noindent
{\it Keywords\/}: counting, satisfiability, complexity, R\'enyi
entropy, entanglement

%\abbreviations{SAT, satisfiability; MPS, matrix product state; CNF,
%  conjunctive normal form; DNF, disjunctive normal form; FPRAS, fully
%  polynomial-time randomized approximation scheme}

%%%%%%%%%%%%%%%%%%%%%%%%%%%%%%%%%%%%%%%%%%%%%%%%%%%%%%%%%%%%%%%%%%%%%%%%%%%%
\section{Introduction}

Satisfiability (SAT) is an NP-complete problem that aims at deciding
whether there is an $n$-bit string input that satisfies a Boolean
logic formula \cite{gavey-johnson,arora-barak}. An example of a
satisfiability problem is Circuit Satisfiability, or CSAT, where a
circuit is built with a number of gates that is polynomial in
$n$. While the cost of testing whether a given $n$-bit string
satisfies the circuit is polynomial, finding whether the circuit is
satisfiable is a hard problem, and counting the number of satisfying
inputs is even harder. The problem of counting satisfying solutions of
SAT is known as \#SAT \cite{valiant,welsh-gale}.

The difference in complexity between finding and counting
solutions~\cite{jerrum-valiant-vazirani} is clearly illustrated in the
case of 2SAT, the problem of satisfiability of logic formulas built
using Boolean expressions involving exactly two literals (or
bits). While the logic expression for a problem in 2SAT can be
determined to be satisfiable or not in polynomial time (2SAT $\in$ P),
it is believed that counting the number of {\it all} satisfying
solutions (when they exist) cannot be done efficiently. Indeed, \#2SAT
is a problem in the class \#P-complete, which is also the same class
containing \#3SAT, although 3SAT is in NP-complete. In other words,
even though it is much easier to solve 2SAT than 3SAT, counting the
satisfying solutions is as difficult in one problem as in the
other. Algorithms for counting exactly the solutions of the \#SAT
problem exist (see for instance \cite{dahlfof05,furer07}), but they
require a number of operations that scales exponentially with $n$. No
polynomial or even subexponential algorithm is known.

The fact that a non-exponential algorithm for counting solutions of a
SAT problem is still unknown raises the following question: Is there
an entropic principle that limits the efficiency of large-scale
counting machines, much as there is one that limits the efficiency of
thermal engines? Here we propose the R\'enyi entanglement entropies as
means to quantify the difficulty of counting problems. We test the
idea explicitly for the case of \#2SAT stated in conjunctive normal
form (CNF) and its negation, which is stated in disjunctive normal
form (DNF).

The entanglement entropy, a concept of much use in quantum information
theory, differs from the thermodynamic entropy. In the case of SAT
problems, the usual entropy $S$ tells us about the number of solutions
$Z$, i.e. $S=\log_2 Z$. For example, if instead of counting the number
solutions one were asked to present all solutions, it would take a
time $O(2^S)$ to do so. But counting the number of solutions is a bit
easier, in the sense that one could ``compress'' the information
needed to do the counting without presenting all solutions. The degree
of compression of the information needed to do the counting is what we
relate below to the R\'enyi entanglement entropies $S^{(q)}$.

In this article, we show rather generically how one can define the
R\'enyi entropies $S^{(q)}$ associated to problems of Boolean
expression satisfiability. We then focus on the particular case of
random \#2SAT problems, compute the R\'enyi entropies $S^{(q\to 0)}$
and $S^{(q=2)}$, and, in certain cases, $S^{(q\to 1)}$ as well.

%%%%%%%%%%%%%%%%%%%%%%%
\begin{table}
\centering
\caption{Summary of the results for the \#2SAT problem.}
%\begin{indented}
\hspace{.5cm}
\begin{small}
\begin{tabular*}{\hsize}{@{\extracolsep{\fill}}ccccccc}
\br
%\cline{1-6}
\multicolumn{3}{c}{CNF} & & \multicolumn{3}{c}{DNF} \\
%\hline
\cline{1-3}\cline{5-7}
R\'enyi &  efficient & conjecture & & R\'enyi & efficient & conjecture  \\
entropy & algorithm & verified? & & entropy & algorithm & verified? \\ 
scaling & known? & & & scaling & known? & \\
%\hline
\cline{1-3}\cline{5-7}
$S^{(0)} \propto n$ &  exact: NO & $\surd$ & & $S^{(0)} \propto n$ & exact NO: & $\surd$ \\
%\hline
\cline{1-3}\cline{5-7}
$S^{(1)} \propto n$ & approximate: &$\surd$ & & $S^{(1)} \rightarrow 0$ & approximate: & $\surd$  \\
$S^{(2)} \propto n$ & NO & & & $S^{(2)} \rightarrow 0$ &YES &  \\ 
%\hline
%\cline{1-3}\cline{5-7}
\br
\end{tabular*}
\end{small}
%\end{indented}
\end{table}
%%%%%%%%%%%%%%%%%%%%%%%

We conjecture that the R\'enyi entropy $S^{(q\to 0)}$ provides
information on the difficulty of counting solutions exactly, while the
R\'enyi entanglement entropies $S^{(q>0)}$ tell us about the
possibility of doing efficient (polynomial-time) approximate
counting. More precisely, we conjecture that what determines if the
exact counting can or cannot be done efficiently is whether $S^{(q\to
  0)}$ scales with $n$ slower or faster than $\log_2 n$. Similarly,
what determines whether an approximate counting can or cannot be done
efficiently is whether $S^{(q>0)}$ scales with $n$ slower or faster
than $\log_2 n$. We then test this conjecture in the case of \#2SAT
and its negated version using the entropies we obtained numerically
for random instances of the problem. We find that in these cases the
entropies give the correct prediction for whether efficient algorithms
should exist or (likely) not. Our results are summarized in Table 1.

%%%%%%%%%%%%%%%%%%%%%%%%%%%%%%%%%%%%%%%%%%%%%%%%%%%%%%%%%%%%%%%%%%%%%%%
\section{R\'enyi entropies and \#SAT}

Let us start by considering a set of $n$ binary variables
$\{x_j=0,1\}_{j=1,\ldots,n}$, and a binary weight $W(x)\equiv
W(x_1,\ldots,x_n)=0$ or $1$, depending on whether the string of binary
variables $x\equiv x_1\,x_2 \ldots x_n$ satisfies or not a given
Boolean expression. With the weights $W(x)$, we define the vector
\begin{equation}
\label{eq:Psi}
|W\rangle = \sum_{x_n,\ldots,x_1=0,1} W(x_1,\ldots,x_n)\, |x_1\ldots
 x_n\rangle,
\end{equation}
where $|x_1\,x_2 \ldots x_n\rangle\equiv |x\rangle$ denotes a
particular configuration of this binary system.

Next, we construct the (unnormalized) density matrix
\begin{equation}
\label{eq:rho}
\varrho
=
|W\rangle \langle W|
= \sum_{x,x'}
W(x)\, W(x')\, 
|x\rangle\langle x'|
\;.
\end{equation}
The number of strings $x$ satisfying the given Boolean expression is
given by
\begin{equation}
\label{eq:Z}
Z
=
\mbox{tr} \;\varrho
=
\sum_{x}
W(x)^2
=
\sum_{x}
W(x)
\;,
\end{equation}
where we used that $W(x)=0,1$. If one wishes, a normalized density
matrix $\rho=\varrho/Z$ can also be constructed.

%%%%%%%%%%%%%%%%%%%%%%%
\begin{figure}
\centering \includegraphics[width=6cm]{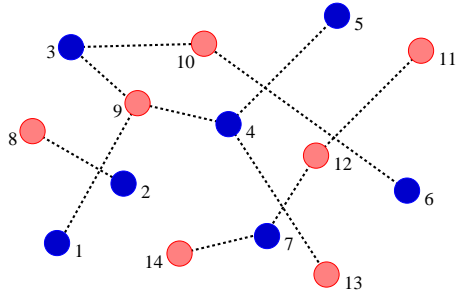}
\caption{Example of a bipartition of a system of 14
  Boolean variables into two sets: $A$ (dark blue) and $B$ (light
  red). The dashed lines represent 2-literal clauses
  entering in the Boolean expression of a 2SAT problem.}
\label{fig1}
\end{figure}
%%%%%%%%%%%%%%%%%%%%%%%

The entanglement R\'enyi entropies are constructed from reduced
density matrices after dividing the system into subsystems $A$ and $B$
(see figure \ref{fig1} for an example). Let system $A$ be comprised of
bits $1$ to $l$, and system $B$ of bits $l+1$ to $n$: $x_A\equiv
x_1x_2\dots x_l$ and $x_B\equiv x_{l+1}\dots x_n$, so that $x\equiv
x_Ax_B$. Next, construct a $2^l\times 2^{n-l}$ matrix ${\cal
  W}_{x_A,x_B}\equiv W(x_Ax_B)$ out of the list of weights $W(x)$. One
then defines the reduced density matrix
\begin{eqnarray}
\label{eq:rho_A}
\varrho_A
&\equiv&
\mbox{tr}_B \;\varrho
\nonumber\\
&=&
\sum_{x^{\,}_A,x'_A} \left(\sum_{x^{\,}_B} W(x^{\,}_Ax^{\,}_B)\,W(x'_Ax^{\,}_B)\right)
|x^{\,}_A\rangle\langle x'_A|
\nonumber\\
&=&
\sum_{x^{\,}_A,x'_A} 
[{\cal W}{\cal W}^\top]_{x^{\,}_A,x'_A}
\;|x^{\,}_A\rangle\langle x'_A|
\;.
\end{eqnarray}
The R\'enyi entanglement entropies \cite{renyi61} are given by
\begin{equation}
\label{eq:Renyi}
S^{(q)}_{AB}
=
\frac{1}{1-q}\,
\log_2\left[
\frac{\mbox{tr}_A\, \varrho_A^q}{(\mbox{tr}_A\, \varrho_A)^q}
\right]
\;.
\end{equation}
It follows from the cyclicity of the trace that
$S^{(q)}_{AB}=S^{(q)}_{BA}$, thus the entropies are independent of the
order of the traces (i.e. of which of $A$ or $B$ is traced out first).

The entanglement entropies depend only on the singular values
resulting from the decomposition ${\cal W}=U\,\Lambda\,V^\top$, where
$U$ is an orthogonal $2^l\times 2^l$ matrix, $V$ is an orthogonal
$2^{n-l}\times 2^{n-l}$ matrix, and $\Lambda$ is a $2^l\times 2^{n-l}$
rectangular diagonal matrix with elements $\lambda_k,
k=1,\dots,d=\min(2^l,2^{n-l})$. The entanglement entropies are given
in terms of these singular values by
\begin{equation}
\label{eq:S-lambda}
S^{(q)}_{AB}
=
\frac{1}{1-q}\,
\log_2\left[
\frac{\sum_{k=1}^d \lambda_k^{2q}}
{(\sum_{k=1}^d \lambda_k^{2})^q}
\right]
\;.
\end{equation}
The number of satisfying solutions $Z$ is also linked to the singular
values:
\begin{equation}
\label{eq:Z-lambda}
Z=
\sum_{k=1}^d \lambda_k^{2}
\;.
\end{equation}

The number of singular values depends on how the system is
partitioned. But the partition into two systems is just one step of
many: one can further recursively split systems $A$ and $B$ each into
two, $A_1$, $A_2$, $B_1$, and $B_2$, and so on. This is a way to
construct a representation of the $W(x)$ as a matrix product state
(MPS). The partitions that lead to the largest number of singular
values are those when systems $A$ and $B$ are of the same order, so we
shall focus on this case for the discussion that follows.

To get the {\it exact} value of $Z$, one must sum over {\it all} the
singular values. There are possibly $d$ of them, but many can be
zero. So how many are there that are non-zero? The number $r$ of
non-zero singular values are related to the R\'enyi entropy with $q\to
0$, which counts all the non-zero $\lambda_k$, all these weighted with
the same factor $\lim_{q\to 0}\lambda_k^{2q}=1$. One finds $\log_2
r=S^{(q\to 0)}_{AB}$.

The rank $r$ sets the size of the matrices that can represent $W(x)$
as a MPS. Therefore, the quantity $S^{(q\to 0)}_{AB}$ is basically a
measure of the amount of resources needed to compress the information
in all the solutions of the SAT problem. One can show~\cite{r-bound}
that $r \le \min(2^l,2^{n-l},Z)$, and thus $S^{(q\to 0)}_{AB}\le
S=\log_2 Z$. Because $S^{(q\to 0)}_{AB}\le S$, the complexity of
counting is smaller than that of listing the solutions, for counting
could in principle be done by working with the compressed information
without the need to expand it.

%%%%%%%%%%%%%%%%%%%%%%%%%%%%%%%%%%%%%%%%%%%%%%%%%%%%%%%%%%%%%%%%%%%%%%%
\section{Numerical computations of $S^{(q)}$ for the \#2SAT problem}

%%%%%%%%%%%%%%%%%%%%%%%
\begin{figure}
\begin{center}
\includegraphics[width=8cm]{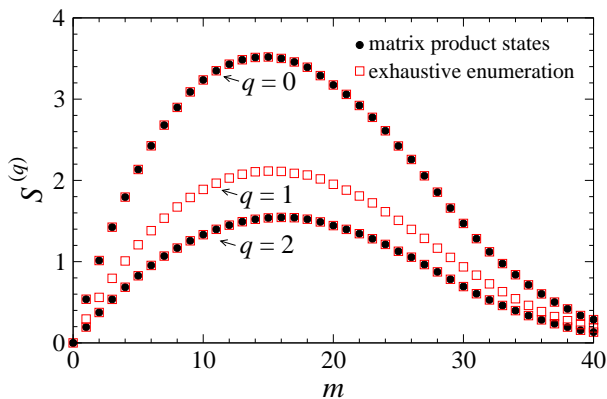}
\end{center}
\caption{R\'enyi entropies for the \#2SAT problem as
  functions of the number of clauses for $n=20$. The data points are
  averaged over 4000 realizations. The graph provides a comparison
  between exhaustive enumeration of solutions and the matrix product
  states method for the cases $q=0$ and $q=2$. The case $q=1$ obtained
  by exhaustive enumeration is also shown.}
\label{fig2}
\end{figure}
%%%%%%%%%%%%%%%%%%%%%%%

Let us henceforth concentrate on calculating the R\'enyi entropies for
the specific case of \#2SAT. We consider random 2SAT problems, where
$m$ clauses are drawn uniformly among $n(n-1)/2$ bit pairs, and among
the four possible clauses that involve an OR and two literals (each of
which can be negated or not). We then compute the entropies $S^{(q\to
  0)}_{AB}$ and $S^{(q=2)}_{AB}$. Data for systems of size up to
$n=20$ can be easily obtained from the weights $W(x)$ by exhaustive
enumeration of all possible inputs $x$ (see figure \ref{fig2});
however, for larger sizes we deploy a matrix computing scheme based on
the method of \cite{matrix-comp} (details of the method are provided
in \ref{app:numerics} and \ref{app:bit-string}). The results from the
matrix computing method for $q=0$ and $q=2$ were compared against the
results of exhaustive enumeration of solutions for the same random
problem, realization by realization, and found to match each other
within the relative error margin controlled by the threshold used to
distinguish the smallest non-zero singular value from zero (typically,
$10^{-10}$). In figure \ref{fig2}, we present results using both
methods for $n=20$ after averaging over 4000 realizations.

\begin{figure}
\begin{center}
\includegraphics[width=8cm]{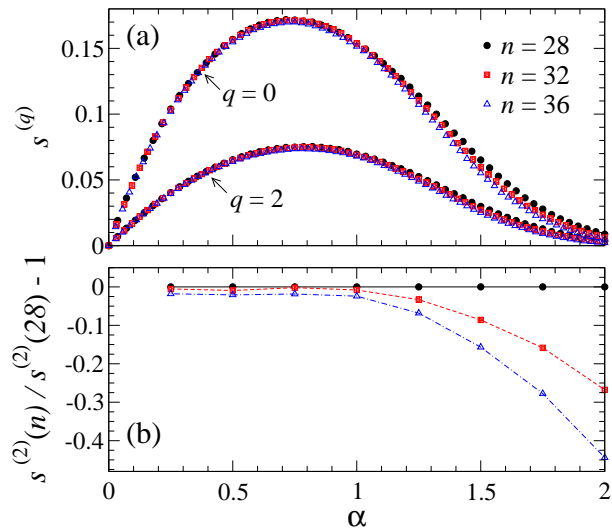}
\end{center}
\caption{(a) Dependence of two R\'enyi entropies per bit, $s^{(0)} =
  S^{(q\rightarrow 0)}/n$ and $s^{(2)} = S^{(q=2)}/n$, with the ratio
  $\alpha=n/m$. The data shows that the entropies per bit are very
  weakly dependent on $n$ for $\alpha<1$. The matrix product state
  method was employed in the simulations and a total of 4000
  realizations were used for each data set. (b) The relative deviation
  $s^{(2)}(n)/s^{(2)}(28)-1$ as a function of $\alpha$, showing in
  more detail the weak dependence of $s^{(2)}$ with $n$ when
  $\alpha<1$. The strong deviation seen when $\alpha>1$ indicates the
  onset of the 2SAT phase transition at $\alpha=1$. The lines are
  guides to the eye.}
\label{fig3}
\end{figure}

In figure \ref{fig3}a we show how the entropies per bit $s^{(0)} =
S^{(q\to 0)}_{AB}/n$ and $s^{(2)} = S^{(q=2)}_{AB}/n$ vary as we
increase the ratio between the number of gates and the number of bits,
$\alpha=m/n$. The numerical data was averaged over 4000 realizations
of random 2SAT problems. The very weak dependence of the entropies per
bit on $n$ lead us to conclude that both $S^{(q\to 0)}_{AB}$ and
$S^{(q=2)}_{AB}$ scale as the ``volume'', i.e. the entropies are very
closely proportional to the system size $n$ up to $\alpha=1$. The
deviation from this linear behavior is very small, less than 3\%, as
shown in figure \ref{fig3}b, and can be attributed to the finite
number of bits in our simulations. A finite-scaling analysis indicates
that the entropies per bit saturate to a finite value as
$n\rightarrow\infty$ when $\alpha<1$. We note that the value
$\alpha=1$ marks the onset of a well-known phase transition in the
random 2SAT problem \cite{goerdt92,chvatal92,monasson96}: for
$\alpha>1$, one expects $S^{(q)}/n \rightarrow 0$ as
$n\rightarrow\infty$ (i.e., the number of solutions goes to zero when
$m>n$). This is consistent with the growing deviation from the linear
behavior that we observe in figure \ref{fig3}b beyond $\alpha=1$.

From the perspective of the matrix computing technique we employed in
our simulations, the difficulty in counting solutions of a given
realization of an $m$-clause \#2SAT problem is not directly related to
the value of $S^{(q)}$ computed after the $m$th clauses. Instead, the
difficulty is measured by the maximum value that $S^{(q)}$ reaches as
the logic gates are applied, and this maximum can be reached at an
intermediate number of clauses $m^\ast$, before all $m$ clauses are
enforced. Typically, R\'enyi entropies reach a maximum value around a
value $m^\ast \approx n$ and quickly decay beyond that point (see
figure \ref{fig2}). Nevertheless, for all values of $m$, the averaged
R\'enyi entropies for CNF always scale linearly with $n$, provided
that $\alpha < 1$.

%%%%%%%%%%%%%%%%%%%%%%%%%%%%%%%%%%%%%%%%%%%%%%%%%%%%%%%%%%%%%%%%%%%%%%%
\section{The negation version of the \#2SAT}

Next, we compare and contrast the entanglement entropies computed for
the 2SAT problems with those for the negation of the same
problems. More precisely, let us consider the Boolean expressions
$\overline W(x)=0$ if $W(x)=1$ and $\overline W(x)=1$ if $W(x)=0$.

\begin{figure}
\begin{center}
\includegraphics[width=7cm]{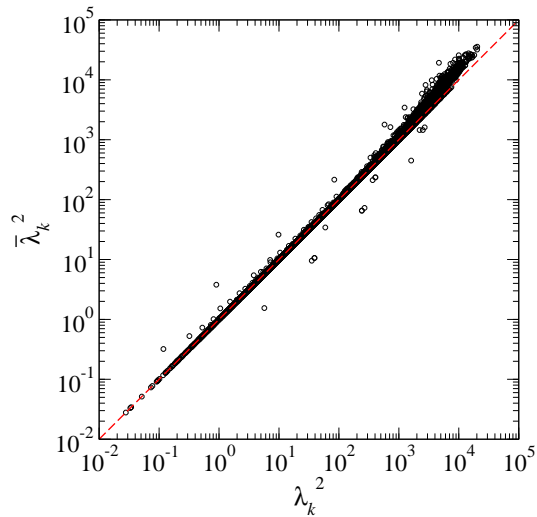}
\end{center}
\caption{Correlation between the singular values associated to $W$ and
  to the negation $\bar W$ for $n=20$. For each of 1000 realizations
  of the problem, the singular values $\lambda^2_k$ for the a random
  2SAT problem and the $\bar \lambda^2_k$ ones associated to the
  negated problem are rank ordered, and the pair
  $(\lambda^2_k,\bar\lambda^2_k)$ is plotted. The total number of
  pairs in the plot, for all the 1000 realizations combined, is
  14827. The dashed straight line is a guide to the eye. The slight
  deviations away from the line at higher values of the singular
  values are dominated by the second largest singular value of the
  negated problem, the one with $k=1$.}
\label{fig4}
\end{figure}

The singular values obtained from the weights $\overline W(x)$ are
intimately correlated to those derived from the weights $ W(x)$. We
find empirically that the relation for large $n$ should be as follows
(see \ref{app:negation} for a heuristic argument) in the case when the
number of satisfying solutions is less than the number of
non-solutions, i.e. $Z<2^n/2$. For each and every realization of the
problem, if there are $r$ non-zero singular values $\lambda^2_k$,
$k=1,\dots,r$, for the 2SAT problem with weights $W(x)$, then there
are $r+1$ singular non-zero singular values $\bar\lambda^2_k$,
$k=0,\dots,r$, for the negated problem with weights $\overline
W(x)$. The relation between these two sets of singular values, which
should hold in the large $n$ limit, is
\numparts
\begin{eqnarray}
\label{eq:negationa}
&&\bar\lambda^2_0=2^n-2Z
\\
\label{eq:negationb}
&&\bar\lambda^2_k=\lambda^2_k, \quad k=1,\dots,r
\;.
\end{eqnarray}
\endnumparts
Notice that $\bar Z= \sum_{k=0}^r \bar\lambda_k^{2}=2^n-Z$, which is
the number of solutions of the negated problem. To support our claim
that the singular values of the two problems are related according to
(\ref{eq:negationa},\ref{eq:negationb}), we plot in figure \ref{fig4}
the pairs of singular values $(\lambda^2_k,\bar\lambda^2_k)$, for
$k=1,\dots,r$. The data displayed are for $n=20$ and 1000 realizations
of the random 2SAT problem and its negation. We also computed the
relative deviation $\epsilon= |\bar\lambda^2_0-(2^n-2Z)|/(2^n-2Z)$ for
the highest singular value, and averaged this deviation over the 1000
realizations. We find that $\epsilon_{\rm ave}<0.3\%$. We conclude
that even for $n$ as small as 20 the relations in
(\ref{eq:negationa},\ref{eq:negationa}) hold rather well.

The relationship between the singular values for $W$ and $\bar W$
yields, via (\ref{eq:S-lambda}), a connection between the respective
R\'enyi entropies:
\begin{equation}
\label{eq:entropy-connection}
2^{(1-q){\bar S}^{(q)}_{AB}}
=
%\left(1-\frac{Z}{\bar Z}\right)^q
%\left(\frac{Z}{\bar Z}\right)^q\;
\left({Z}/{\bar Z}\right)^q\;
2^{(1-q){S}^{(q)}_{AB}}
+\left(1-{Z}/{\bar Z}\right)^q
\;.
\end{equation}
In particular
\begin{eqnarray}
\label{eq:entropy-connection-0}
2^{{\bar S}^{(q\to 0)}_{AB}}
=
2^{{S}^{(q\to 0)}_{AB}}+1
\;,
\end{eqnarray}
as expected since there is one more singular value in the negated
problem, and
\begin{equation}
\label{eq:entropy-connection-1}
\fl {\bar S}^{(q\to 1)}_{AB}
=
\left(\frac{Z}{\bar Z}\right)\,
{S}^{(q\to 1)}_{AB}
%\\
%&&\!\!\!\!\!\!\!\!\!\!\!\!\!\!\!\!\!\!
- \left[
%\left(1-{Z}/{\bar Z}\right)\,\log_2\left(1-{Z}/{\bar Z}\right)
%+\left({Z}/{\bar Z}\right)\,\log_2\left({Z}/{\bar Z}\right)
\left(\frac{Z}{\bar Z}\right)\,\log_2\left(\frac{Z}{\bar Z}\right)
+\left(1-\frac{Z}{\bar Z}\right)\,\log_2\left(1-\frac{Z}{\bar Z}\right)
\right]
\;.
%\nonumber
\end{equation}

For a \#2SAT problem with a non-zero ratio $\alpha$ of clauses to
variables, the number of satisfying solutions is exponentially smaller
than the number of unsatisfying solutions: $\lim_{n\to \infty} Z/\bar
Z=0$. Therefore, we find that in this case
%$\lim_{n\to \infty}{\bar S}^{(q\to 1)}_{AB} =0$.
${\bar S}^{(q>1)}_{AB}\to 0$ in the large $n$ limit.

%%%%%%%%%%%%%%%%%%%%%%%%%%%%%%%%%%%%%%%%%%%%%%%%%%%%%%%%%%%%%%%%%%%%%%%
\section{R\'enyi entropies and the complexity of \#SAT}

With these results for the R\'enyi entropies, we are in a position to
test whether or not, at least for the \#2SAT and its negation, the
entropies can predict the degree of difficult in solving a problem.

$\bullet$ {\it 2SAT solutions} -- In the case of \#2SAT, we find that
both $S^{(q\to 0)}$ and $S^{(q=2)}$ are volumetric, i.e. they scale
linearly with $n$. Since $S^{(q\to 0)} \ge S^{(q\to 1)}\ge S^{(q=2)}$,
the entanglement entropy with $q\to 1$ is also volumetric. Because
$S^{(q\to 0)}$ is volumetric, carrying out the counting exactly cannot
be done efficiently, which is expected since, after all, \#2SAT is
\#P-complete and hence unavoidably hard. Thus, in this case, the
entropy is a good predictor of the difficult of the problem. Moreover,
because $S^{(q\to 1)}$ is also volumetric, we do not expect that an
approximate counting algorithm should exist either \cite{welsh-gale}.

$\bullet$ {\it 2SAT non-solutions} -- In the case of the negation of
\#2SAT, we find that ${\bar S}^{(q\to 0)}$ is volumetric, while ${\bar
  S}^{(q\to 1)}$ vanishes for large $n$. We would then conclude that
counting the exact number of non-satisfying binary strings is still
unavoidably hard. However, because ${\bar S}^{(q\to 1)}\to 0$, one can
efficiently approximate the number of non-satisfying inputs. This
results is in agreement with what is known about the problem, as we
explain below in more detail.

That counting exactly in one problem is equivalent to counting exactly
in the other is evident, as $\bar Z=2^n-Z$. Thus, if one finds $Z$ one
has $\bar Z$ and vice versa. But since $\bar Z\gg Z$, it is easier to
find an approximation to $\bar Z$ than to $Z$ within the same {\it
  relative} error. The issue can be alternatively stated as follows. A
2SAT problem is presented in conjunctive normal form (CNF), with $m$
OR-clauses with two literals, all AND-ed together. Its negation is
then in disjunctive normal form (DNF), with $m$ AND-clauses with two
literals, all OR-ed together. It is known that DNF problems admit a
fully polynomial-time randomized approximation scheme
(FPRAS)~\cite{vazirani-book}. That the entropy ${\bar S}^{(q\to 1)}\to
0$ while, in contrast, ${S}^{(q\to 1)}$ is volumetric, is the way in
which our entanglement entropy approach signals that the DNF problem
is simpler than the CNF one.

%%%%%%%%%%%%%%%%%%%%%%%%%%%%%%%%%%%%%%%%%%%%%%%%%%%%%%%%%%%%%%%%%%%%%%%%
\section{Conclusion}

Let us conclude with a few remarks. First, for CNF problems defined on
a graph, one may expect that the entanglement should be
volumetric. However, it is not always the case that a problem defined
on graph will have volumetric entanglement, as exemplified in the DNF
problem that results from the negation of a CNF one. The intuition as
to why that is so is that the disjunctive form splits the problems
into many disjoint ones, whereas the conjunctive ones ties the bits
together, and is mean-field like. The entanglement is one way to
capture the fact that some problems cannot be simplified by dividing
into subproblems, while others can.

Finally, we would like to mention a possible practical way to explore
the entanglement as predictor of whether it may or not be possible to
approximately count for a given problem. Notice that we succeeded in
getting the scaling of the R\'enyi entropies even for systems of
modest sizes. Therefore, one could get a sense of whether a FPRAS may
possibly exist, without explicitly constructing one, using an
entanglement entropy finite-size scaling analysis.

%%%%%%%%%%%%%%%%%%%%%%%%%%%%%%%%%%%%%%%%%%%%%%%%%%%%%%%%%%%%%%%%%%%%%%%%
\ack
This work was supported in part by the NSF grants CCF-1116590 and
CCF-1117241. We would like to thank H. Castillo, C. Laumann, and
P. Wocjan for useful discussions.

\appendix

%%%%%%%%%%%%%%%%%%%%%%%%%%%%%%%%%%%%%%
\section{Numerical simulations with matrix product states}
\label{app:numerics}

We employed the matrix computing method introduced in
\cite{matrix-comp} to verify numerically that $S^{(q)} \propto n$ for
the conjunctive form of the \#2SAT problem. In this method, we
associate to each binary variable $x_j$ a pair of real matrices
$M_j^{0}$ and $M_j^{1}$ of dimensions $D_{j-1} \times D_{j}$. The
weight $W(x)$ of each configuration $x$ is written as the trace of a
product of these matrices, namely,
\begin{equation}
\label{eq:P}
W(x_1,\ldots,x_n) = \mbox{tr} \left( M_1^{x_1} \cdots
M_n^{x_n} \right).
\end{equation}
(The trace can be dropped if we consider the first and last matrices
to be row and column vectors, respectively, namely, $D_0 = D_n = 1$.)
It is straightforward to show that $Z$, the number of satisfying
configurations, is given by the expression
\begin{equation}
Z = \mbox{tr} \left[ \left(M_1^0 + M_1^1 \right) \cdots \left( M_n^0 +
  M_n^1 \right) \right].
\end{equation}
The matrix representation of the weight $W(x)$ allows us treat
(\ref{eq:Psi}) similarly to a quantum mechanical superposition
state. Then, by operating sequentially on adjacent pairs of matrices
$(M_j^{x_j},M_{j+1}^{x_{j+1}})$, we can simultaneously check the
satisfiability of all $2^n$ instances of a CNF problem.

One starts by setting $W=W_0(x)=1$ for all values of the $n$-bit
string $x$. This can be easily implemented by choosing $D_j=1$ and
$M_j^0=M_j^1=1$ for all $j=1,\ldots,n$, yielding $Z=Z_0=2^n$. Each
clause $C_k$ in a CNF can eliminate non-satisfying instances of the
problem and, consequently, reduce $Z$. We call these gate operations
filters. Thus, starting from an initial weight distribution $W_0(x)$,
the vector $|W_0\rangle$ evolves into a state $|W_C\rangle$ by the
sequential application of filters that block states that do not
satisfy a CNF Boolean expression $C=C_1 \wedge C_2 \cdots \wedge
C_m$. Notice that there are four possible types of two-bit OR filter
gates, depending on whether the input bits are negated or not.

Since the order in which bits appear in the clauses in $C$ is random
and only adjacent bit operations are allowed in matrix computing,
matrices have to be moved up and down the bit string. This is done
through SWAP gates \cite{matrix-comp}. These SWAP gates, when combined
with other logic gates such as OR, tend to rapidly increase the rank
of the matrices. This is a limiting factor of the method, as every
gate operation between bits $(j-1,j)$ employs a singular value
decomposition which requires $O(D_{j-1}^3)$ floating point
operations. On the other hand, the filters, by state elimination,
partially compensate the growth in matrix rank and $D_{\rm max} = {\rm
  max}\{D_j\}$ tend to peak around $m\approx n$. Therefore, for each
realization of a CNF, the computational cost of the numerical
calculation scales as $O(D_{\rm max}^3)$.

In order to minimize the number of SWAP operations and limit the
growth of matrix ranks, we employ two pre-processing
strategies. First, we reorder the bits in the string so that the sum
of pair-wise distances between bits in the clauses of the CNF is
minimized. This is done through a Monte Carlo sampling with a 100\%
rejection rate if the new configuration has a higher total distance
than the previous one. Typically, 30 sampling steps are used.

Second, we reorder the clauses so that those bits participating in few
or no clauses are acted on first. Bits which are no longer required
are set to an inactive state by absorbing their bit matrices into
matrices of a neighboring bit and replacing their matrices with
identity ones. These measures reduce the matrix ranks substantially,
which in turn allow us to count exactly all solutions for large
CNFs. We find empirically that $\langle D_{\rm max}^3 \rangle \sim
2^{0.1 n}$. This favorable scaling (as compared to the computational
cost of the best known algorithm for solving exactly the \#2SAT
problem, which scales as $2^{0.329n}$ \cite{furer07}) corresponds to
an average behavior and does not apply to hard, worst-case
realizations of the CNF.

%%%%%%%%%%%%%%%%%%%%%%%%%%%%%%%%%%%%%%
\section{Bit-string partitioning}
\label{app:bit-string}

Once the complete sequence of $m$ clauses of a CNF is implemented with
matrix computing, the partition matrix ${\cal W}$ can in principle be
obtained in the following way. Consider the bit-string matrix set
$\{M_j^{x_j}\}$ resulting from the sequence of Boolean
gates. Following the definition of ${\cal W}$, we can write
\begin{equation}
\label{eq:Wdecomp}
{\cal W}_{x_A; x_B} = \sum_{\alpha=1}^{D_{n/2}} A_{x_A;\alpha}\,
B_{\alpha;x_B},
\end{equation}
where
%
%\numparts
\begin{eqnarray}
A_{x_1\cdots x_{n/2};\alpha} & = & M_1^{x_1} \cdots M_{n/2}^{x_{n/2}} \\
B_{\alpha;x_{n/2+1}\cdots x_n} & = & M_{n/2+1}^{x_{n/2+1}} \cdots
M_n^{x_n}.
\end{eqnarray}
%\endnumparts
%
This decomposition of ${\cal W}$ can be used to compute some R\'enyi
entropies without the need to obtain ${\cal W}$ explicitly. While it
is straightforward to show that $S^{(q\rightarrow 0)} = \log_2
D_{n/2}$, $S^{(q>1)}$ requires an additional manipulation. Combining
the singular value decomposition ${\cal W} = U\, \Lambda\, V^\top$ and
with (\ref{eq:Wdecomp}), we can write
\begin{equation}
\label{eq:BBAA}
\sum_{k=1}^{D_{n/2}} \lambda_k^{2q} = {\rm tr} \left( \Lambda^{2q}
\right) = {\rm tr} \left[ \left(U^\top {\cal W} {\cal W}^\top U
  \right)^q \right] = {\rm tr} \left[ \left( B B^\top A^\top A
  \right)^q \right],
\end{equation}
where we used the cyclicity of the trace. Finally, combining
(\ref{eq:S-lambda}) with (\ref{eq:BBAA}), one finds $S^{(q>1)}$.

The two major advantages in this approach as compared to performing
the singular values decomposition of ${\cal W}$ is the reduced number
of floating point operations (since no singular value decomposition of
${\cal W}$ is needed) and the reduced memory allocation for matrix
storage. The disadvantage is that one cannot compute $S^{(q\rightarrow
  1)}$. In practice, this is the only viable numerical approach that
we know for finding the scaling of $S^{(q)}$ with $n$ when $n>20$.

%%%%%%%%%%%%%%%%%%%%%%%%%%%%%%%%%%%%%%
\section{Heuristic argument for (\ref{eq:negationa},\ref{eq:negationb})}
\label{app:negation}

Because $\overline W(x)=1-W(x)$, the associated matrix $\overline{\cal
  W}={\cal M}-{\cal W}$, where ${\cal W}$ is the $2^l\times 2^{n-l}$
matrix constructed in the main text and ${\cal M}$ is a matrix of the
same size with all entries equal to 1. It follows that
\begin{eqnarray}
\overline{\cal W}\, \overline{\cal W}^\top &=& {\cal M}{\cal
  M}^\top-{\cal W}{\cal M}-{\cal M}{\cal W}^\top+{\cal W}{\cal W}^\top
\nonumber\\ &=& 2^{n-l}\, {\cal H}-{\cal W}{\cal M}-{\cal M}{\cal
  W}^\top+{\cal W}{\cal W}^\top \;,
\end{eqnarray}
where we defined the $2^l\times 2^l$ matrix ${\cal H}$ with all
entries equal to 1. Now,
\begin{equation}
\fl \left[{\cal W}{\cal M}+{\cal M}{\cal W}^\top\right]_{ij}
%&=&
%\sum_{k=1}^{2^{n-l}} {\cal W}_{ik}+\sum_{k=1}^{2^{n-l}} {\cal W}^\top_{kj}
%\nonumber\\
= \sum_{k=1}^{2^{n-l}} {\cal W}_{ik}+\sum_{k=1}^{2^{n-l}} {\cal
  W}_{jk} \approx
\frac{1}{2^l}\,2\,\sum_{q=1}^{2^{l}}\sum_{k=1}^{2^{n-l}} {\cal W}_{qk}
\approx \frac{1}{2^l}\,2 Z\;,
\end{equation}
where we used that the average over the lines $i$ and $j$ of the
matrix ${\cal W}$ should become independent of the line index for
large enough matrices. We thus have that
\begin{eqnarray}
\overline{\cal W}\, \overline{\cal W}^\top
&\approx&
(2^{n}-2Z)\,\frac{1}{2^l} {\cal H}+{\cal W}{\cal W}^\top
\;.
\end{eqnarray}
The matrix ${\cal H}$ has one eigenvalue equal to $2^l$, and $2^l-1$
zero eigenvalues. Therefore, $(2^{n}-2Z)\,\frac{1}{2^l} {\cal H}$
alone has one eigenvalue $\lambda_0^2=2^{n}-2Z$, and $2^l-1$ zero
eigenvalues. Let us discuss the situation when $Z<2^n/2$, in which
case $\lambda_0^2=2^{n}-2Z>0$. One can then add ${\cal W}{\cal
  W}^\top$ as a perturbation, which lifts up the massive degeneracy
without affecting much the non-zero eigenvalue because of the large
splitting. Therefore, the non-zero eigenvalues of $\overline{\cal W}\,
\overline{\cal W}^\top$ should be all the non-zero eigenvalues
$\lambda^2_k$ of ${\cal W}{\cal W}^\top$, plus the extra large
$\lambda_0^2$ eigenvalue. This is the heuristic argument for
(\ref{eq:negationa},\ref{eq:negationb}), which we support numerically
as explained in the text.

We remark that we arrived at the result that the negated problem with
weights $\overline W(x)=1-W(x)$ has one more singular value than the
problem with weights $W(x)$ using that the number of solutions of
$W(x)=1$ is less than the number of non-solutions $W(x)=0$,
i.e. $Z<2^n/2$. This ensured that the $\lambda_0^2=2^{n}-2Z$ singular
value was positive; if it were negative, the perturbative argument
should break down, for one would have to restore the positiveness of
all the singular values in the end. Therefore what determines which of
the two problems, with weights $W(x)$ or $\overline W(x)$, has more
singular values is which one has more solutions. That we used
$Z<2^n/2$ breaks the symmetry between the two problems, and explains
why we cannot use the argument twice doing a double negation
$\overline{\overline W}(x)$. The argument only goes in one direction,
and applies only to the case when $Z<\bar Z$.

%%%%%%%%%%%%%%%%%%%%%%%%%%%%%%%%%%%%%%%%%%%%%%%%%%%%%%%%%%%%%%%%%%%%%%%%%%%

%%%%%%%%%%%%%%%%%%%%%%%%%%%%%%%%%%%%%%%%%%%%%%%%%%%%%%%%%%%%%%%%


\begin{thebibliography}{15}
\bibitem{r-bound} There are in total $Z$ non-vanishing matrix elements
  in ${\cal W}$. There are also at most $Z$ lines in ${\cal W}$ that
  do not contain only 0s (the upper bound is reached if there are $Z$
  linearly independent lines, each line with a single entry that is
  equal to 1). Hence, $r \le \min (Z,2^l)$. An analogous argument for
  the columns yields $r\le \min (Z,2^{n-l})$. Therefore, $r\le \min
  (2^l,2^{n-l},Z)$.
\end{thebibliography}
\end{document}